\documentclass[amsmath,amssymb,epsfig,pre,aps]{revtex4}

\usepackage{graphicx}
\usepackage{dcolumn}
\usepackage{bm}

\begin{document}

\title{Traveling solitons in the damped
driven nonlinear Schr\"odinger equation}
\author{I.V. Barashenkov$^\dag$}
\author{E.V. Zemlyanaya$^\S$}
\affiliation{Department of Applied Mathematics,
University of Cape
Town, Rondebosch 7701, South Africa}


\begin{abstract}
The well known effect of the linear damping on the moving
nonlinear Schr\"odinger soliton
(even when there is a supply of energy  via the spatially homogeneous driving)
is to quench its momentum to zero. Surprisingly, the zero
momentum does not necessarily mean zero velocity. We show
that two or more parametrically driven damped solitons
can form a complex  traveling with zero momentum
at a nonzero constant speed.

All traveling complexes we have found so far, turned out to
be unstable. Thus, the parametric driving is capable of
sustaining the uniform motion of damped solitons, but some 
additional agent is required to stabilize it.
\end{abstract}

\maketitle

\section{Introduction}
The amplitude of a nearly-harmonic wave propagating in a nonlinear
dispersive medium satisfies a nonlinear Schr\"odinger
 equation.
Confining ourselves to the generic, cubic,
 nonlinearity of the `focusing'
type,
the resulting nonlinear Schr\"odinger equation is of the form
\begin{equation}
i \Psi_t +\Psi_{xx} +2 |\Psi|^2 \Psi =-i \gamma \Psi;
\quad \gamma>0.
\label{0}
\end{equation}
The $-i \gamma \Psi$ term in the right-hand side accounts for
dissipative losses (which were assumed to be small in the derivation
of eq.(\ref{0}).) In the underlying
physical system the dissipation is compensated by pumping the
energy into the system, in one way or another. The pumping is modeled
by adding a driving term to the right-hand side of eq.(\ref{0}).

Like a simple pendulum, the distributed system can
be driven externally or parametrically. The typical form of
the corresponding amplitude equation is
\begin{equation}
 i \Psi_t +\Psi_{xx} +2 |\Psi|^2 \Psi =
 h e^{i\Omega  t} -i \gamma \Psi,
\label{direct}
\end{equation}
and
\begin{equation}
\label{eq1}
 i \Psi_t +\Psi_{xx} +2 |\Psi|^2 \Psi =
 h {\overline \Psi} e^{2i\Omega  t} -i \gamma \Psi,
\end{equation}
respectively.
(The overline in  the right-hand side of 
(\ref{eq1}) indicates complex conjugation.)
Both the externally and parametrically driven nonlinear
Schr\"odinger equations arise in a great variety of physical contexts.
In particular, the parametric equation (\ref{eq1}) 
describes the nonlinear Faraday
resonance in a vertically oscillating water tank
\cite{Miles,ElphickMeron1,Faraday}
and the effect of phase-sensitive 
amplifiers on solitons in optical fibers \cite{optics}.
The same equation controls the magnetization
waves in an easy-plane ferromagnet exposed to a combination
of a static and microwave field 
\cite{BBK} and the amplitude of synchronized oscillations 
in vertically vibrated pendula lattices \cite{pendula}.

Both  equations (\ref{direct}) and (\ref{eq1})
exhibit soliton solutions
\cite{RoySoc,Smirnov,boundary}, \cite{Miles,ElphickMeron1,BBK},
stable and unstable \cite{Smirnov,BBK}, which can also form
(stable and unstable)
multisoliton complexes \cite{Malomed,BSA,BZ}.
All localized solutions that have been found so far, 
were non-propagating. In fact, it is widely
accepted that the nonlinear Schr\"odinger
 solitons simply {\it cannot\/} travel in the presence
of dissipation. This perception is based on 
the  rate equation
\begin{equation}
\label{eq5}
\dot P = -2\gamma P,
\end{equation}
 which is straightforward from
(\ref{direct}) and (\ref{eq1}). Here $P$ is the total field momentum,
\begin{equation}
\label{eq3} P=\frac i 2 \int_{-\infty}^{\infty} (\overline \Psi_x \Psi
- \Psi_x \overline \Psi) dx.
\end{equation}
 In the undamped case ($\gamma=0$)
the momentum is conserved; however if $\gamma >0$, $P$ decays
to zero and this seems to suggest that a
 solitary wave, initially moving with a nonzero velocity,
 will have to
 slow down and eventually stop \cite{KN_PRB}.

Another indication that only quiescent solitons are possible in the
damped-driven  Schr\"odinger equation, 
comes ostensibly from the singular 
\cite{ElphickMeron1,ElphickMeron2}
and
Inverse Scattering-based perturbation
theory \cite{RoySoc,KivsharMalomed,Shesnovich}. Here we should mention however that
these techniques are well developed only in the one-soliton
sector and in the 
case of several well separated solitons.
They either
 make use of the smallness of the perturbation
in the right-hand side of (\ref{direct})-(\ref{eq1}) \cite{ElphickMeron1,RoySoc,KivsharMalomed}
or utilize an explicit form of the perturbed soliton (to study
its stability and bifurcation) \cite{Shesnovich}.
In any case, the resulting finite-dimensional system of
equations for the parameters of the soliton and
radiations, leads to the conclusion that
the soliton's velocity has to decay to zero as $t \to \infty$.

Meanwhile, the moving solitary waves
  could play a significant role
in a variety of physical situations modeled by the
damped-driven nonlinear
Schr\"odinger equations. Stable traveling waves could compete
with non-propagating localized attractors; unstable ones might arise as
long-lived transients and intermediate states in
spatiotemporal chaotic regimes.  Both types of moving
solitary waves
could mediate energy dissipation in damped-driven systems.
One more reason for not rejecting the unstable
solutions outright is  their possible persistence within the (directly
or parametrically driven)
Ginzburg-Landau equations of which the  Schr\"odinger
equations (\ref{direct})-(\ref{eq1}) are special cases
\cite{Ginzburg_Landau}. 
The diffusion
and nonlinear damping 
(the  terms
$ic_1 \Psi_{xx}$ and $-ic_2|\Psi|^{2n} \Psi$, to be added to
 the right-hand side of
(\ref{direct})-(\ref{eq1}))
are known to have a stabilizing effect on the Ginzburg-Landau
pulses; hence the unstable Schr\"odinger solitons may gain
stability as they are continued to nonzero positive $c_1$
and $c_2$.

The purpose of this paper is to
show that the damped-driven nonlinear Schr\"odinger equations do support
solitary waves traveling with nonzero velocities.
For the demonstration of this fact
 we confine our study to the {\it parametrically\/}
driven Schr\"odinger  only. The {\it externally\/} driven equation
 is left as an object of future research.

Two complementary strategies will be pursued to
achieve our goal. First,
in section \ref{nonzero_gamma},
we consider the {\it motionless damped\/} solitons
($V=0$, $\gamma \neq 0$)
and derive the condition under which they can be continued
to nonzero velocity. Having identified
values of $\gamma$ for which this condition is satisfied,
we perform the numerical continuation
obtaining a branch of solitary waves with nonzero $V$ and $\gamma$.
Our second approach is presented in
section \ref{nonzero_V}; the idea
is to continue {\it undamped traveling\/} waves
($\gamma=0$, $V \neq 0$)
to nonzero dampings. We show that this is only possible
if the traveling wave has zero momentum.
For complexes with $P=0$,  we then carry out the numerical
continuation in $\gamma$.
In section \ref{consistency} we discuss the consistency of results
obtained within these two complementary approaches.

We examined, numerically, stability of all solutions obtained within both
approaches. 
The general framework of the
stability analysis is outlined
in section \ref{Preliminaries}. Results of this analysis are presented
along with
results of the numerical continuation.
Finally, section \ref{conclusions} summarizes conclusions of our study.

\section{Mathematical Preliminaries}
\label{Preliminaries} 

For purposes of this paper we
transform equation (\ref{eq1})
to an autonomous form.
First, we normalize the driving frequency $\Omega$ to
unity; after that, the substitution $\Psi (x,t) = e^{it}\psi (x,t)$ takes
eq.(\ref{eq1}) to
\begin{equation}
\label{eq2}
 i \psi_t +\psi_{xx} +2 |\psi|^2 \psi  -\psi = h \overline \psi
-i \gamma \psi.
\end{equation}
This is the representation of the
parametrically driven damped nonlinear Schr\"odinger
equation that we are going to work with in this paper.
We confine ourselves to uniformly traveling
solutions of the form
\begin{equation}
\label{eq7}
 \psi(x,t)=\psi(x-Vt)\equiv \psi(\xi),
\end{equation}
 where $\psi(\xi) \rightarrow 0$
 as $|\xi| \rightarrow \infty$. These satisfy an ordinary differential
equation
\begin{equation}
\label{eq8}
-iV \psi_{\xi} +\psi_{\xi\xi} + 2 |\psi|^2 \psi -\psi
= h \overline \psi-i\gamma\psi.
\end{equation}

The analytical part of this paper deals mainly with identifying
those of the previously found solutions of (\ref{eq8})
 with $V=0$ or $\gamma=0$
which can be continued in $V$ and $\gamma$, respectively. The
actual continuation will be  carried out numerically. Our
numerical method employs a predictor-corrector continuation
algorithm with a fourth-order accurate Newtonian solver.
Typically, the infinite line was approximated by an interval
$(-100,100)$.  The discretization step size was typically $0.005$.
The numerical tolerance was set to be $10^{-10}$; that is, the
grid solution would be deemed accurate if the difference between
the left- and right-hand sides in (\ref{eq8}) were smaller than
$10^{-10}$.

Along with the continuation of solutions in $V$
and $\gamma$, we will be
analyzing their stability to small
perturbations. To this end, we linearize equation eq.(\ref{eq2})
in the co-moving frame of reference.
Letting $\psi(x,t)= u(\xi)+iv(\xi)+\delta \psi (\xi,t)$,
where $u$ and $v$ are the real and imaginary part of the 
solution  that we are linearizing about,
and assuming that the
linear perturbation  depends on time exponentially:
$$\delta \psi (\xi,t) =  e^{\lambda t}
\left[ \delta u (\xi) + i\delta v(\xi) \right],
$$
we arrive at an eigenvalue problem
\begin{equation}
\label{eq25}
{\cal H}_0 \, {\vec y} = (\lambda + \gamma)J \, {\vec y},
\end{equation}
where the   operator ${\cal H}_0$ is defined by
\begin{equation}
\label{H0}
 {\cal H}_0 = \left(
 \begin{array}{lr}
 -\partial_\xi^2+1+h -6u^2-2v^2 & -V \partial_\xi -4uv \\
 V \partial_\xi -4uv & -\partial_\xi^2+1-h -6v^2-2u^2
 \end{array} \right),
\end{equation}
the skew-symmetric matrix $J$ is
\[
J = \left(
\begin{array}{lr} 0 & -1 \\
                  1  &  0
\end{array}
\right),
\]
and the column vector
${\vec y}(\xi)= (\delta u,\delta v)^T$.
The eigenvalue problem  (\ref{eq25}) was solved by expanding $\delta u$
and $\delta v$ over a Fourier basis, typically with 500 modes,
on the interval $(-50,50)$.

The last point that we need to touch upon in this preliminary
section, is the integrals of motion of eq.(\ref{eq1}), or,
more precisely, the quantities which are conserved in the
absence of  dissipation.
When $\gamma=0$, the equation (\ref{eq1}) conserves the momentum
(given by eq.(\ref{eq3}) where one only needs to replace
$\Psi \to \psi$),
and energy,
\begin{equation}
\label{eq4}
E= \int_{-\infty}^{\infty} (|\psi_x|^2+ |\psi|^2 -
|\psi|^4+h\ {\rm Re} \, \psi^2) \, dx.
\end{equation}
In the damped case,  the momentum decays according to the rate
equation (\ref{eq5})
while the energy satisfies
\begin{equation}
\label{eq6}
\dot E = 2\gamma \left ( \int_{-\infty}^{\infty}
|\psi|^4dx-E \right ).
\end{equation}


\section{Continuation of damped solitons
to nonzero velocities}
\label{nonzero_gamma}
\subsection{Continuability criterion}

Our first strategy is to attempt to continue
stationary solutions with nonzero
$\gamma$
to nonzero $V$.
Two basic soliton solutions, denoted
 $\psi_+$ and $\psi_-$, are available explicitly:
\begin{eqnarray}
\psi_\pm(x)=e^{-i \theta_\pm} A_\pm \, {\rm sech\/}
\left( A_\pm x \right);
 \label{plus_minus}
\\
A_\pm= \sqrt{ 1\pm \sqrt{h^2-\gamma^2} },
\nonumber
\\
\theta_+= \frac12 \arcsin \frac{\gamma}{h},
\quad \theta_-=\frac{\pi}{2}-
\theta_+.
\nonumber
\end{eqnarray}
The two solitons can form a variety of  stationary complexes.
These are denoted, symbolically,  $\psi_{(++)}$, $\psi_{(--)}$,
$\psi_{(+-+)}$, $\psi_{(-+-)}$,
and so on \cite{BZ}.
Let $\psi_0(x)$ be a particular complex; we want to find out
whether it can be continued in $V$. Assuming there is a solution
$\psi(\xi;V)$ such that $\psi(\xi;0) \equiv \psi_0(\xi)$
$(=\psi_0(x))$,
we expand $\psi(\xi;V)$
in powers of $V$ as
\begin{equation}
\left. \psi(\xi;V)= e^{-i \theta} \right\{
u_0(\xi)+iv_0(\xi) 
\left.+ V[u_1(\xi)+iv_1(\xi)]
  +V^2[u_2(\xi)+iv_2(\xi)]+ ...
\right\},
\label{expansion}
\end{equation}
where the constant phase $\theta$ will be chosen at a later stage. 
We also expand
$h$ and $\gamma$: $h=h_0+h_1V+...$,
$\gamma= \gamma_0+\gamma_1V+...$.
Substituting into (\ref{eq8}),
the order $V^1$ gives
\begin{equation}
\label{u1v1}
 {\cal L} \left(
\begin{array}{l}
u_1  \\ v_1
\end{array} \right) =\left( \begin{array}{c}
v_0' \\ -u_0'
 \end{array} \right)+
{\cal B} \left( \begin{array}{c}
u_0 \\ v_0
 \end{array} \right),
\end{equation}
where the operator ${\cal L}$ has the form
\begin{equation}
\label{L}
 {\cal L} =
 \left(
 \begin{array}{lr}
 -\partial_x^2+1+h_0 \cos 2 \theta-6u_0^2-2v_0^2 & 
 \gamma_0 + h_0 \sin 2 \theta -4u_0v_0\\
 -\gamma_0+ h_0 \sin 2\theta  -4u_0v_0   & 
  -\partial_x^2+1-h_0 \cos 2 \theta-2u_0^2-6v_0^2
 \end{array} \right);
\end{equation}
the constant matrix ${\cal B}$ is given by
\[
{\cal B}=\left ( \begin{array}{lr}
-h_1 \cos 2 \theta &  -\gamma_1-h_1 \sin 2\theta \\
 \gamma_1-h_1 \sin 2\theta & h_1 \cos 2 \theta
\end{array} \right),
\]
and the primes over $u_0$ and $v_0$ indicate  derivatives with
respect to $x$. (In (\ref{u1v1}) and (\ref{L}) we have
replaced $\xi$ with $x$ as $\xi$ coincides with
$x$ for $V=0$.)
According to
Fredholm's alternative, eq.(\ref{u1v1}) has a bounded
solution $u_1(x), v_1(x)$ iff
the vector in the right-hand side is orthogonal to the kernel of the
Hermitean-conjugate operator ${\cal L}^\dagger$:
\begin{equation}
\int (y,w) \, \left( \begin{array}{c}
v_0' \\ -u_0'
 \end{array} \right) dx + \int (y,w) \, {\cal B} \left( \begin{array}{c}
u_0 \\ v_0
 \end{array} \right)d x=0.
\label{continuability}
\end{equation}
Here ${\vec y}(x)=(y,w)^T$ is the eigenvector of ${\cal L}^\dagger$
associated with the zero eigenvalue: ${\cal L}^\dagger {\vec y}=0$.
That the operator ${\cal L}^\dagger$ has a zero eigenvalue follows from
the fact that the operator
${\cal L}$ has one --- namely, the translation
eigenvalue corresponding to the eigenvector $(u_0', v_0')^T$.
Eq.(\ref{continuability}) gives a necessary continuability
condition of damped
quiescent solitons to nonzero velocities.

\subsection{Noncontinuability of the `building blocks'}
\label{Nonconti}

It is quite easy to check that when
$\gamma_0 \neq 0$, the individual $\psi_+$ and $\psi_-$
solitons
(the basic `building blocks'  of
which all complexes are constructed)
are {\it not\/} continuable to nonzero $V$.
Choosing $\theta=\theta_+$ for the $\psi_+$ soliton
and $\theta=\theta_-$ for the  $\psi_-$
(where $\theta_\pm$ are to be computed from
the bottom formula in (\ref{plus_minus}) with $\gamma=\gamma_0$
and $h=h_0$),  we get $v_0(x)=0$,
 $\gamma_0-h_0 \sin2 \theta=0$ and
so the $2 \times 2$ matrix ${\cal L}$,  eq.(\ref{L}),
 becomes upper triangular.
The zero mode
of ${\cal L}^\dagger$ can now be readily found.

Consider, for instance,
the $\psi^+$ case. The zero mode satisfies
\[
\left(
\begin{array}{lr}
-\partial_x^2+A_+^2-6u_0^2 & 0 \\
2 \gamma_0 & -\partial_x^2+A_-^2-2u_0^2
\end{array}
\right)
\left( \begin{array}{c}
y \\ w
 \end{array} \right)=0,
 \]
hence
$y(x)=u_0'(x)$ and $w(x)$ is found from
\begin{equation}
(-\partial_x^2+ A_-^2-2u_0^2)w=- 2 \gamma u_0'(x).
\label{zm}
\end{equation}
Using the explicit expression for $u_0(x)$, $u_0(x)=A_+ {\rm sech\/}(A_+x)$,
the operator in the left-hand side of (\ref{zm}) can be written
as $A_+^2(L_0-\epsilon)$, where $\epsilon=2h_0 \cos(2
\theta_+)/A_+^2$; $L_0$ is given by
\[
L_0=
-\partial_X^2+ 1- 2 {\rm sech\/}^2 X,
\]
and $X=A_+x$.
 The operator $L_0$ has familiar
spectral properties; in particular it has a single discrete
eigenvalue $E_0=0$ associated with an
even eigenfunction $z_0={\rm sech\/} X$, while its continuous
spectrum occupies the semi-axis $E_k \geq 1$. Consequently,
for $0< \epsilon <1$ (that is, for $h_0 < \sqrt{1+\gamma_0^2}$),
the operator $L_0-\epsilon$ is invertible and  a bounded
solution $w(x)$ of (\ref{zm}) exists and is unique.
It can be found explicitly, but this is not really necessary for our
purposes. All we need to know is that, since $L_0$ is
 a parity-preserving operator,  $w(x)$ has the same
parity as the right-hand side in (\ref{zm}), i.e. it is an odd function.
For that reason the second integral in
equation (\ref{continuability}) vanishes
and the necessary continuability
condition reduces to
\begin{equation}
\gamma \int u_0'(x) (L_0-\epsilon)^{-1} u_0'(x) dx=0.
\label{continua}
\end{equation}
This quadratic form can be easily evaluated by expanding $u_0'(x)$
over eigenfunctions of the operator $L_0$:
\[
u_0'(x)= \int_{-\infty}^{\infty} U(k) z_k(X) dk,
\]
where $L_0 z_k(X)=(1+k^2)z_k(X)$.
(The `discrete' eigenfunction $z_0(X)$ does not appear in the expansion
as it has the opposite parity to $u_0'(x)$.) 
Utilizing the orthonormality of the eigenfunctions,
the continuability condition (\ref{continua}) is transformed into
\begin{equation}
\gamma \int \frac{|U(k)|^2} {k^2+1-\epsilon} dk=0.
\label{con_plus}
\end{equation}
As $\epsilon<1$, this condition can obviously not be satisfied
(unless $\gamma=0$).

In the case of the $\psi_-$ soliton the analysis is similar. In
this case the continuability condition (\ref{con_plus}) is
replaced by
\[
\gamma \int \frac{|U(k)|^2} {k^2 +(1-\epsilon)^{-1}} dk=0,
\]
and this cannot be met for the same reason as eq.(\ref{con_plus}).

\subsection{Continuation of the complexes}

\begin{figure}
\includegraphics[ height = 2in, width = 0.5\linewidth]{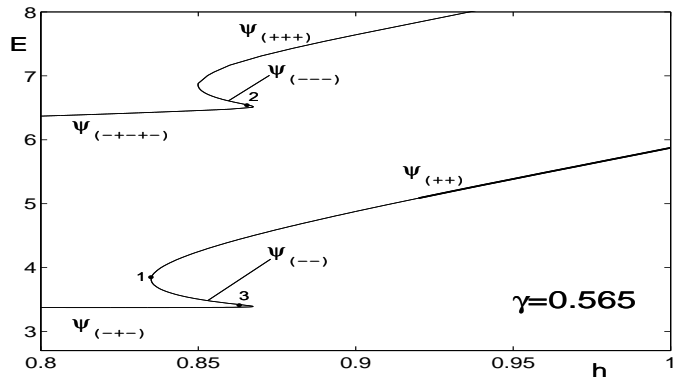}
\caption{\sf A
fragment of the bifurcation diagram for
 stationary multisoliton complexes (adapted from \cite{BZ}.)
 Shown is the energy (\ref{eq4})
 of the complex as a function of $h$.
 The bottom branch pertains to symmetric two-soliton
 complexes  $\psi_{(++)}$ and $\psi_{(--)}$
 and a three-soliton
 solution  $\psi_{(-+-)}$;
 the top branch includes the three-soliton states
 $\psi_{(+++)}$ and $\psi_{(---)}$, as well as a five-soliton
 solution $\psi_{(-+-+-)}$.
 The thick curve corresponds to stable and thin curves to
 unstable solutions.
 The black dots indicate points where the integral (\ref{Ih})
 equals zero and therefore moving solitons are allowed to bifurcate
 off.
 }
\label{E_of_h}
\end{figure}

Turning to the {\it complexes\/} of the solitons
$\psi_+$ and $\psi_-$, the phase of the
complex
varies with $x$ and therefore the matrix ${\cal L}$
cannot be made triangular no matter how we choose the constant
$\theta$ in (\ref{expansion}). For this reason,
aggravated by the fact that the multisoliton solutions are not
available explicitly, the continuability condition
(\ref{continuability}) cannot be verified analytically.
Resorting to the help of computer, we evaluated
the eigenfunction ${\vec y}(x)$
associated with the zero eigenvalue
of the operator ${\cal L}^\dagger$
numerically. (Here we set $\theta=\theta_+$).

All
damped soliton complexes found in \cite{BZ}, were symmetric;
that is, the corresponding $u$ and $v$ are {\it even\/} functions
of $x$.
 Therefore, the operator ${\cal L}^\dagger$ whose potential part
is made up of $u(x)$ and $v(x)$, is
parity preserving and  all its eigenfunctions pertaining to 
non-repeated eigenvalues are either even or odd.
As we move along a continuous branch of
solutions, the parity of the eigenfunction has to change continuously.
Since the parity equals either $+1$ (for even functions)
or $-1$ (for odd functions), the only opportunity left to it
by the continuity argument, is to remain constant on the
entire branch. For that reason it is sufficient to determine the parity
of the eigenfunction for one specific value of $h$ --- and we will know
it at all other points. Our numerical calculation shows that the
eigenfunction ${\vec y}(x)$ is {\it odd\/}
on all branches reported in \cite{BZ}. Consequently, the second term in
(\ref{continuability}) is always zero and we only need to evaluate
the first term.

The vanishing of the term involving coefficients $h_1$ and
$\gamma_1$ in eq.(\ref{continuability}) implies that it
was not really necessary to expand
$h$ and $\gamma$ in powers of $V$. 
This fact has a simple geometric interpretation. 
As we 
will see below, for the fixed $\gamma$ the continuable solutions 
occur only at isolated values of $h$; hence they exist only 
for $h$ and $\gamma$ lying on 
continuous curves in the $(h, \gamma)$-plane.
Each curve results from  an intersection
 of some surface in the 
three-dimensional $(h, \gamma, V)$-space and the  $V=0$-plane.
  The fact that 
one does not have to alter $h$ and $\gamma$ when continuing the 
solution to nonzero $V$, indicates that these surfaces are orthogonal
to the $V=0$ plane along  their curves of intersection.

Having found the solution $\psi(x)=u(x)+iv(x)$ at representative points
along each branch, we obtained the eigenfunction ${\vec y}(x)$
at these points and evaluated
what remains of the integral (\ref{continuability}):
\begin{equation}
\int (y v_0' -w u_0') \, dx \equiv I(h).
\label{Ih}
\end{equation}
The integral $I$ is a continuous function of $h$, and it was
not difficult to find points on the curve at which it
changes from positive to negative values, or vice versa.

\begin{figure}
\includegraphics[ height = 2in, width = 0.5\linewidth]{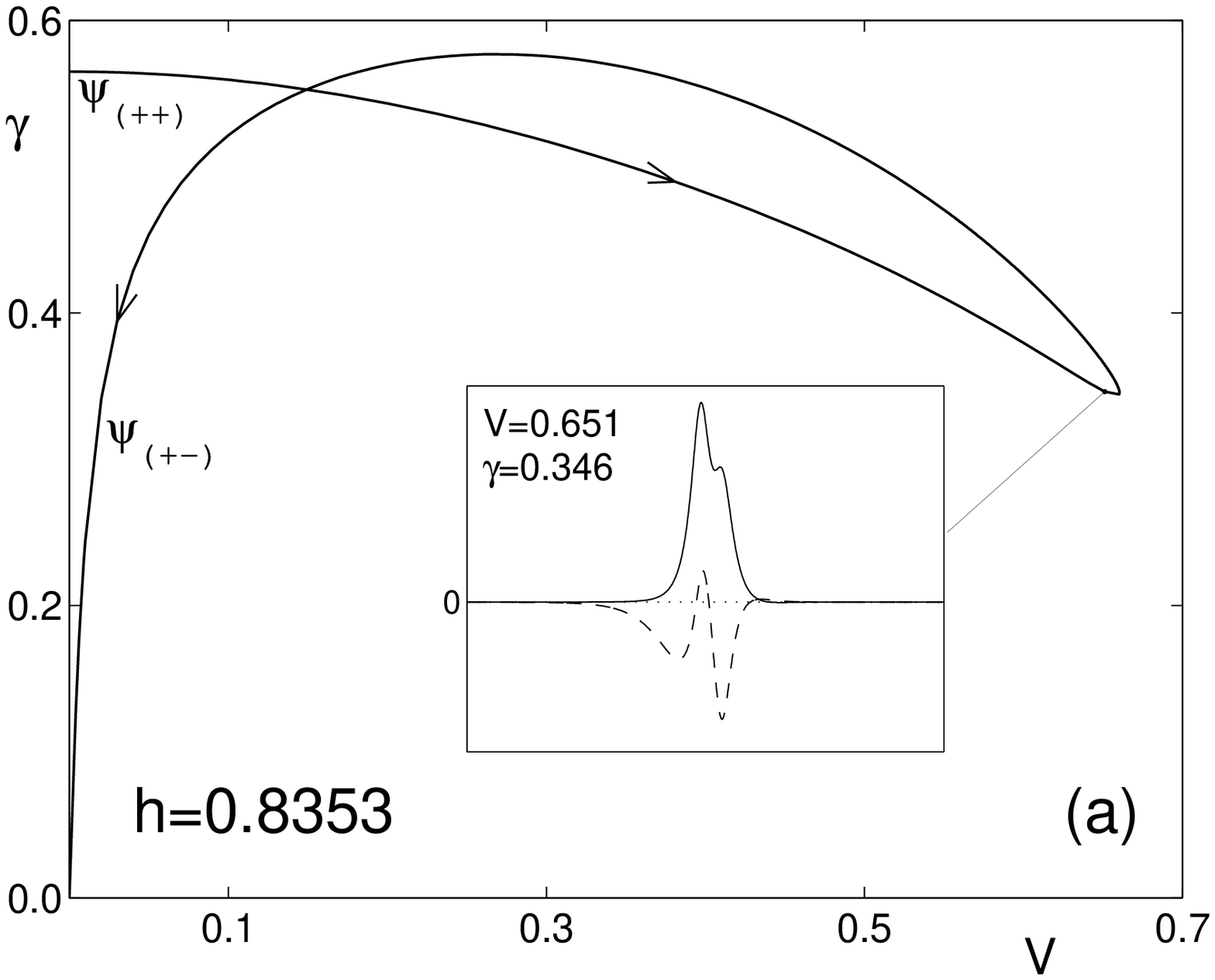}
\includegraphics[ height = 2in, width = .5\linewidth]{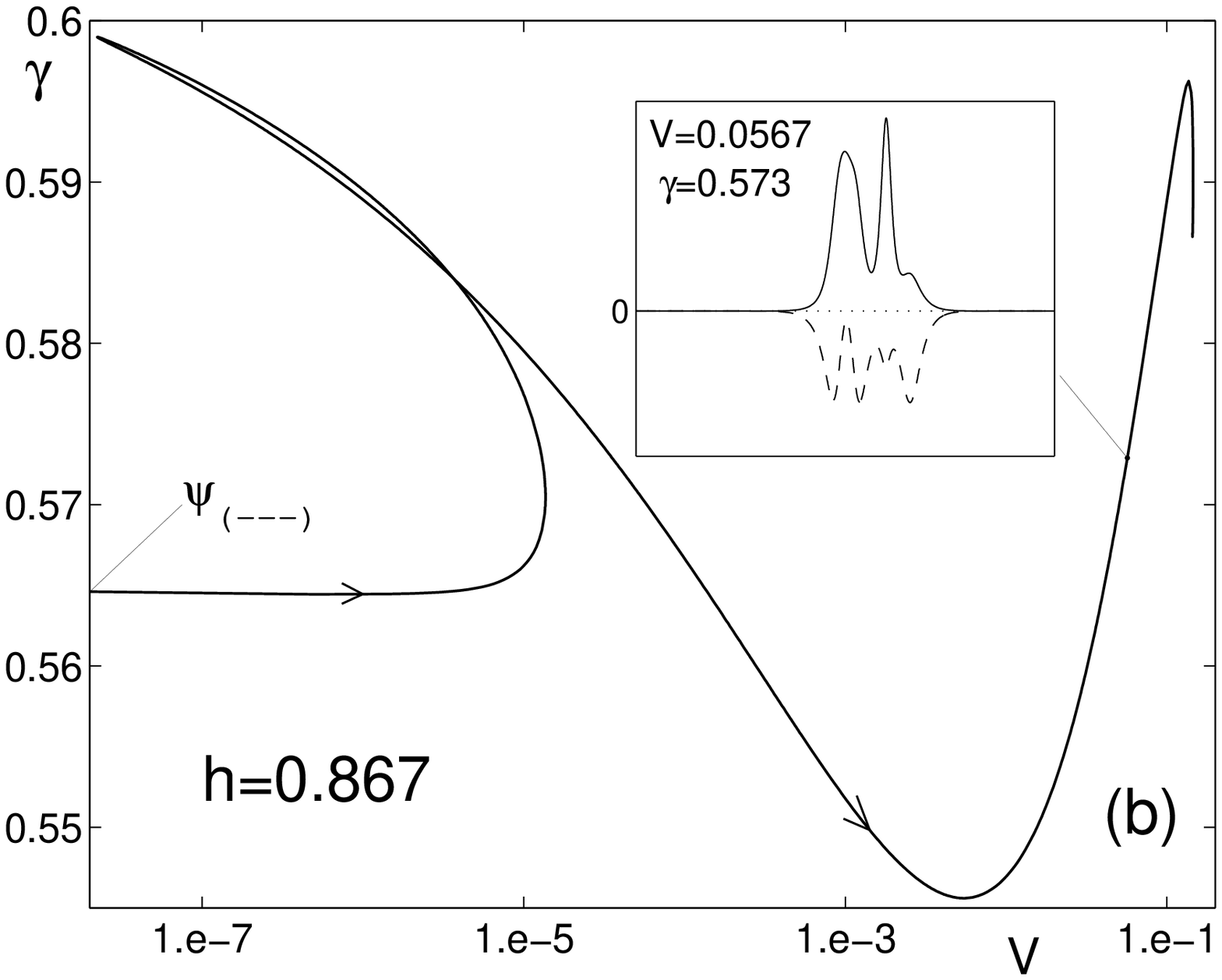}
\includegraphics[ height = 2in, width = 0.5\linewidth]{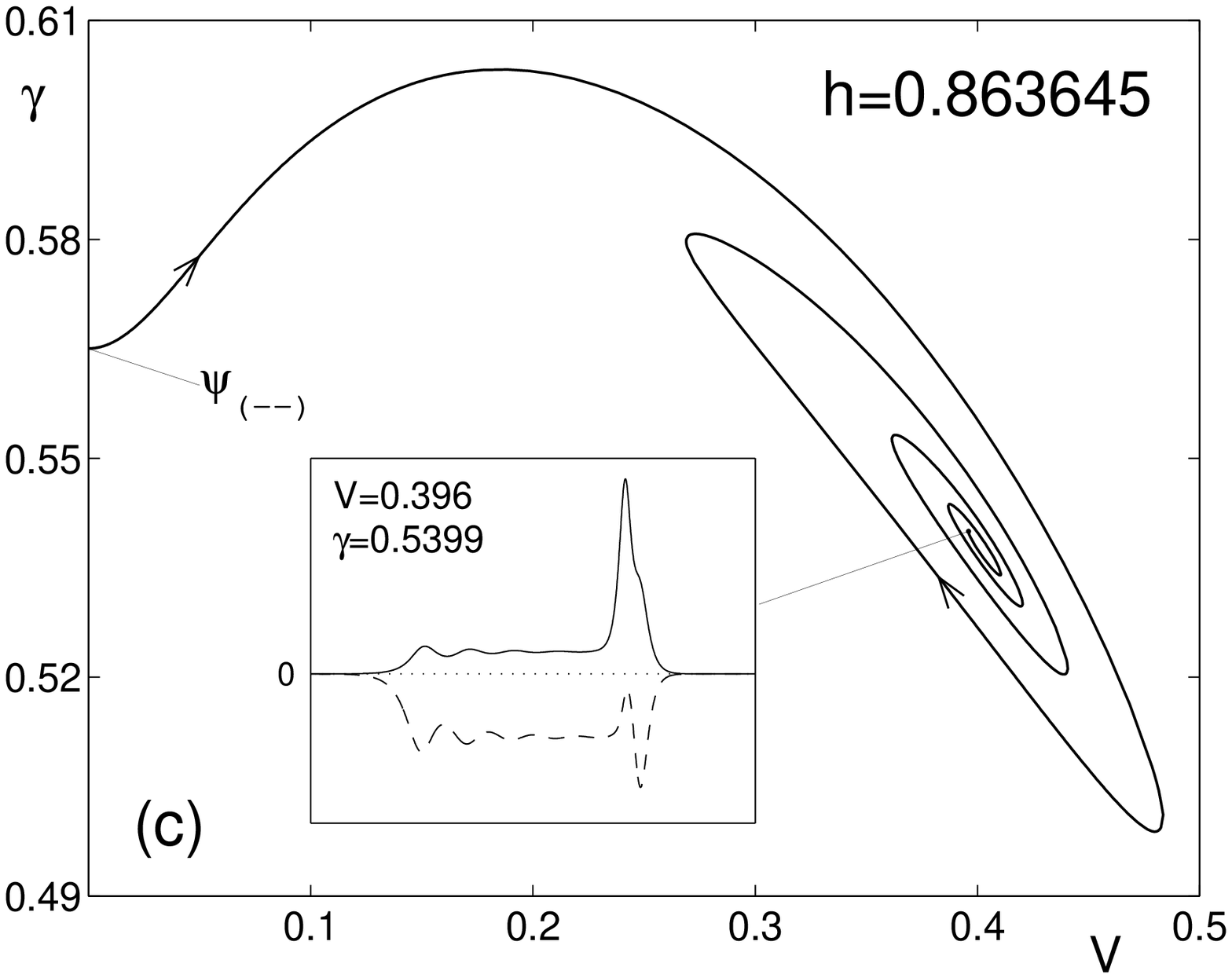}
\caption{\sf Bifurcation curves branching off
the points marked by black dots in Fig.\ref{E_of_h}. The curves
illustrate the
relation between the value of the damping $\gamma$ and velocity $V$
at which the
wave may travel for that  $\gamma$.
Each curve begins at the point $\gamma=0.565$ on the vertical 
axis.
The insets
show representative solutions at internal points of
each branch. (Solid line: real part; dashed line: imaginary
part.)
Note the logarithmic scale of $V$ in (b).
Here, and in all other diagrams, arrows indicate our direction
of continuation.
 }
\label{conti}
\end{figure}

We examined two branches of
multisoliton solutions obtained previously \cite{BZ} (Fig.\ref{E_of_h}).
 The integral $I(h)$ was found to change its sign at three
points, marked by black dots in Fig.\ref{E_of_h}.
(Although it may seem from the figure that $I$ equals
zero right at the turning points, in the actual fact 
zeros of $I$ do not {\it exactly\/} coincide with the turning
points.)
 We were indeed
able to numerically
continue our solutions in $V$ from each of these three points.
Results are presented in Fig.\ref{conti}, (a)-(c).

The point `1' in Fig.\ref{E_of_h} corresponds to the 
stationary complex
$\psi_{(++)}$ and lies just above the turning point
where the $\psi_{(++)}$ turns into $\psi_{(--)}$. (The turning point
has $h=0.83504217$ while $I(h)=0$ for $h=0.8353$.)
 This solution has four positive real eigenvalues
 in the spectrum of the associated linearized operator
and hence is unstable.
 The  $\gamma(V)$
curve which results from the continuation of this solution 
 in $V$, is shown in Fig.\ref{conti}(a). 
As $V$ grows from zero, the solution looses its even symmetry 
(see the inset to Fig.\ref{conti}(a)) while
the four positive
eigenvalues collide, pairwise, and become two complex conjugate
pairs with positive real parts.
 After reaching a maximum velocity of
approximately 0.65, the curve turns back toward $V=0$, with
$\gamma$ first growing but then also turning
toward $\gamma=0$. The solution transforms into a
(strongly overlapped) $\psi_{(+-)}$ complex. As $V$ and $\gamma$
tend to their zero values, the separation between the $\psi_+$ and
$\psi_-$ constituent solitons in the complex grows to infinity.
The spectrum becomes the union of the eigenvalues of the 
individual $\psi_+$
and $\psi_-$ solitons; in particular, it includes a
complex-conjugate pair with positive real part, and a positive
real eigenvalue. Thus the entire branch shown in Fig.\ref{conti}(a)
is unstable.

One more comment that we need to make here, concerns
the validity of the continuation scenario presented in
Fig.\ref{conti}(a) for other values of $h$.
Note  that if we chose a smaller 
value of $\gamma$ in Fig.\ref{E_of_h}, the value 
of $h$ corresponding to the point `1' would also be smaller.
(For example, for $\gamma=0.548$ the integral
$I(h)$ vanishes at the point $h=0.82$.) For this
smaller $h$ the  final product of the continuation turns out to 
be not a pair of infinitely 
separated stationary $\psi_+$ and $\psi_-$ but a totally different
complex. This is discussed below in section \ref{consistency};
see also Figs.\ref{gamma_V}(c) and \ref{cusps}. 

Another branch bifurcates off at the point marked `2' in Fig.\ref{E_of_h}.
Here $h=0.867$. The corresponding $\gamma(V)$ diagram is displayed
in Fig.\ref{conti}(b). As we move along the branch departing from
$V=0$, the original stationary
complex $\psi_{(---)}$ transforms into a solution
displaying three widely separated peaks in its real part:
one corresponding to a strongly overlapping complex $\psi_{(-+-)}$;
the next one to the $\psi_+$ and the last one to the $\psi_-$ soliton. After
passing a
turning point, the curve is reapproaching, tangentially,
the $V=0$-axis.
 However, having reached $V=2.2 \times 10^{-8}$,
it suddenly turns back and the velocity starts to grow
again. The separation
between the solitons decreases and the solution can now be interpreted
as a strongly
overlapping four-soliton complex $\psi_{(+++-)}$
(shown in the inset to Fig.\ref{conti}(b)). As we continue further, the
four constituent solitons regroup into two complexes,
$\psi_{(++)}$ and $\psi_{(+-)}$.
 The distance between the two complexes
   grows rapidly and, for certain finite
 $V$ and $\gamma$  (at the endpoint of the
curve in Fig.\ref{conti}(b)) becomes infinite.
At this point we have two coexisting solutions,
$\psi_{(++)}$ and  $\psi_{(+-)}$, and so this
point corresponds to  the 
 point of self-intersection of the curve shown in Fig.\ref{conti}(a).
 Continuing the two solutions, separately, from the endpoint
 of the curve in Fig.\ref{conti}(b), we reproduce the diagram
 of Fig.\ref{conti}(a)
 for a slightly different value of $h$ (i.e. for $h=0.867$.)
 
 The entire branch shown in
Fig.\ref{conti} (b) is unstable. The start-off stationary solution
$\psi_{(---)}$ has three positive real eigenvalues in its
spectrum; one of these persists for all $V$ and $\gamma$ while the
other two collide and form a complex-conjugate pair with a positive
real part.

The branch continuing from
 the point `3' in Fig.\ref{E_of_h}, for which
$h=0.863645$, leads to the least expected solutions. The
resulting $\gamma(V)$ curve is shown in Fig.\ref{conti}(c). 
For points lying on the 
`spiral' part of the curve, the function $\psi(x)$ is equal to a constant
in a 
relatively large but finite region, and zero outside that region. 
(See the inset to Fig.\ref{conti}(c).)
The constant is $\psi^{(0)}=(A_-/\sqrt{2})e^{-i\theta_-}$;
it defines a stationary spatially uniform 
  solution to
eq.(\ref{eq2}). (This flat background is 
unstable with respect to the continuous
spectrum perturbations.
Figuratively 
speaking, our
pulse solution $\psi(x)$ represents a `droplet' of the unstable phase in the 
stable one.)
On one side (at the rear of the pulse)
 the zero background is connected to the background
$\psi^{(0)}$ by a kink-like interface.
 In the front of the pulse, the interface has the character of 
a large-amplitude excitation, with the shape reminding the 
$\psi_{(+-)}$ complex. As the curve $\gamma(V)$ spirals toward
its `focus' in Fig.\ref{conti}(c), the length of the region
where $\psi(x)=\psi^{(0)}$  is growing. The entire
branch is unstable; the start-off $\psi_{(--)}$ solution already
has two real positive eigenvalues in its spectrum and more appear
as we move along the branch. Those additional positive eigenvalues
are remnants of the `unstable' interval of the continuous spectrum of
the flat nonzero solution $\psi^{(0)}$.

\section{Continuation of traveling waves
to nonzero dampings}
\label{nonzero_V}
\subsection{Continuability conditions}
\label{condition}

When $\gamma=0$, the  equation (\ref{eq8})  has a plethora of localized
solutions with  nonzero $V$ \cite{Baer}, and our second strategy will be
 to attempt to
continue these undamped traveling waves to
nonzero $\gamma$. We start with establishing the necessary and
sufficient conditions for such a continuation.

A set of the {\it necessary\/} conditions can be  easily derived
using two integral characteristics of equation (\ref{eq2}), the
momentum
\begin{equation}
P=(i/2) \int (\overline{\psi}_x \psi-
\psi_x \overline{\psi})dx,
\label{P_psi}
\end{equation}
 and energy (\ref{eq4}).
No matter whether
$\gamma$ equals zero or not, the uniformly
traveling solitary waves (i.e. solutions of the form (\ref{eq7}))
satisfy $\dot P = \dot E =0$. Using these relations 
in eqs.(\ref{eq5})
and (\ref{eq6})
with $\gamma\neq 0$, we get
\begin{equation}
\label{eq9} P=0,
\end{equation}
and
\begin{equation}
\label{eq10}
E=\int|\psi|^4 dx.
\end{equation}
 Equations (\ref{eq9})-(\ref{eq10}) have to be satisfied
by the undamped solutions continuable to nonzero $\gamma$.

In fact, eqs.(\ref{eq9}) and (\ref{eq10}) are not independent.
Indeed, multiplying Eq.(\ref{eq8}) by $\overline \psi$, adding its complex
conjugate and integrating, gives an identity
\begin{equation}
\label{eq11} E-\int|\psi|^4 dx = VP.
\end{equation}
 Letting $P=0$ in
(\ref{eq11}),  eq.({\ref{eq10}) immediately follows.
Thus we can keep  equation
$P=0$ as the {\it only\/}
 necessary condition for the continuability
to nonzero $\gamma$;
eq.(\ref{eq10}) is satisfied as soon as eq.(\ref{eq9}) is in place.

 It turns out that $P=0$ is also a {\it sufficient\/}
condition.
To show this, we expand the field $\psi=u+iv$ in powers of
$\gamma$:
\[
 u=u_0+\gamma u_1+\gamma^2 u_2+..., \quad
 v=v_0+\gamma v_1+\gamma^2 v_2+...,
\]
substitute into (\ref{eq8}) and equate coefficients of like
powers.
(We could have also expanded $h$ and $V$ in $\gamma$, but,
similarly to the continuation in $V$ described
in the previous section, the terms with coefficients $h_1$ and $V_1$ 
cancel out of the resulting continuability condition.)
At the order ${\cal O}(\gamma^1)$, we obtain:
\begin{equation}
\label{eq18} 
{\cal H}_0  \left(
\begin{array}{l}
u_1  \\ v_1
\end{array} \right) =
\left ( \begin{array}{r}
 -v_0\\
u_0 \end{array} \right ).
\end{equation}
Here the hermitian  operator ${\cal H}_0$ is as in (\ref{H0})
where we only need to attach zero subscripts to $u$ and $v$:
\begin{equation}
\label{eq17}
 {\cal H}_0 = \left(
 \begin{array}{lr}
 -\partial_\xi^2+1+h -6u_0^2-2v_0^2& -V \partial_\xi -4u_0v_0\\
 V \partial_\xi -4 u_0v_0 & -\partial_\xi^2+1-h-2u_0^2-6v_0^2
 \end{array} \right).
\end{equation}
Since the operator ${\cal H}_0$ has a zero eigenvalue, with the
translation mode as an
associated eigenvector,
equation (\ref{eq18}) is only solvable if its right-hand side is
orthogonal to $(u'_0,v'_0)$:
\begin{equation}
 \int (u'_0,v'_0)
\left ( \begin{array}{r} -v_0 \\
u_0 \end{array} \right ) \, d\xi \phantom{=}
   = 0.
\label{P=0}
\end{equation}
(Here the prime indicates the derivative with respect to $\xi$.)
The expression in the left-hand side of (\ref{P=0}) 
coincides with equation (\ref{P_psi})
written in terms of the real 
and imaginary part of $\psi$, and so the solvability
condition (\ref{P=0}) is simply $P=0$.

Now assume that $P$ is equal to zero so that a bounded solution  to 
equation (\ref{eq18}) exists.
All traveling waves found in \cite{Baer} have even real and
odd imaginary parts: $u_0(-x)=u_0(x)$, $v_0(-x)=-v_0(x)$.
Noticing that the diagonal elements of the operator ${\cal H}_0$
are parity-preserving while the off-diagonal elements
change their sign under the  $\xi \to -\xi$ reflection, we conclude 
that $u_1(x)$ is odd and $v_1(x)$ is even.

 Proceeding to 
the order ${\cal O}(\gamma^2)$, we have
\begin{equation}
\label{u2v2} 
{\cal H}_0  \left(
\begin{array}{l}
u_2  \\ v_2
\end{array} \right) =
\left ( \begin{array}{r}
 -v_1 +u_0[6u_1^2+2v_1^2]+4v_0 u_1v_1\\
u_1 +v_0[2u_1^2+6v_1^2] +4u_0 u_1v_1
 \end{array} \right ).
\end{equation}
The top entry in the right-hand side of (\ref{u2v2})
is even and the bottom one odd; hence the right-hand side
is orthogonal to the null vector $(u_0',v_0')$ and a bounded 
solution $u_2(\xi), v_2(\xi)$ exists. 
This time the $u$-component is even and the $v$-component odd:
$u_2(-\xi)=u_2(\xi)$,  $v_2(-\xi)=-v_2(\xi)$.

It is not difficult to verify
that this parity alternation 
property guarantees the boundedness of 
$u_n(\xi)$ and $v_n(\xi)$  for all $n$.  
Therefore, equation (\ref{eq8}) 
 has a  localized solution ($\psi(\xi) \to 0$ as 
 $|\xi| \to \infty$) for sufficiently small $\gamma$.
Thus
if we have an undamped soliton traveling with zero
momentum, it can be continued to nonzero values of $\gamma$.

\subsection{Continuable solutions:
the bifurcation diagram of the undamped nonlinear Schr\"odinger}

In this subsection we review 
the $P(V)$ dependence for the
undamped solitons and solitonic complexes \cite{Baer}. Of
interest, of course, are  points where the graph crosses the
$V$-axis, i.e. where $P(V) = 0$.

The simplest solutions arising for
$V=0$ are, obviously, our stationary fundamental solitons
$\psi_+$ and $\psi_-$. These are  given by eqs.(\ref{plus_minus})
where one only needs to set $\gamma=0$:
\[
\psi_+ (x) = A_+ {\rm sech} \, (A_+ x), \quad
\psi_- (x) = iA_- {\rm sech} \, (A_- x),
\]
with $A^2_{\pm}= 1\pm h$. Both $\psi_+$ and $\psi_-$ have zero
momenta  and therefore, are continuable to nonzero
$\gamma$. However, the continuation does not produce any
traveling waves in this case; all we get is our static damped
solitons $\psi_{\pm}$,
 eq.(\ref{plus_minus}).

Next, both $\psi_+$ and $\psi_-$ admit continuation to nonzero $V$ (for
the fixed $\gamma=0$) \cite{Baer}. As $V$ is increased to
\[
c=\sqrt{2+2\sqrt{1-h^2}},
\]
the width of the soliton $\psi_-$ increases, its amplitude
decreases and the soliton gradually transforms into the trivial
 solution, $\psi\equiv 0$. On the resulting branch, the
  momentum vanishes only
for $V=0$ and $V=c$ and therefore, no damped branches can
bifurcate off the traveling $\psi_-$ soliton.

We now turn to the soliton $\psi_+$.
When $h<0.28$, its fate  is similar to that
of the $\psi_-$: as $V\rightarrow c$, the soliton spreads out and
merges with the zero solution. The momentum equals zero only at two
points, $V=0$ and $V=c$; for $0<V<c$, the momentum is positive.

For $h>0.28$, the transformation of the $\psi_+$ is more
promising from the present viewpoint (see the dashed
curve in Fig.\ref{P_of_V}). As $V$ is increased from zero, the momentum
grows, then  the branch turns back toward the $V=0$ axis.
For some $V<0$ the momentum reaches its maximum and then
decreases to zero. The point $V=V_1$ where $P(V_1)=0$ is of
interest to us as a branch of damped solitons can bifurcate off
at this point (and it really does, see subsection
\ref{Large}.)
Continuing beyond $V_1$, the curve $P(V)$ turns
toward $V=0$ and then, after one more turning point, we have
another zero crossing: $P(V_2)=0$. This is how far we have managed
to advance in our previous work \cite{Baer}. 

At this point we need to mention that
the $\psi_+$ and $\psi_-$ are not the only quiescent
solitons for $\gamma=0$. The dashed $P(V)$ curve
in Fig.\ref{P_of_V}  is seen
to have one more intersection with the $P$
axis, apart from the one at the origin.
 The corresponding solution represents a symmetric strongly overlapping
 complex
of the $\psi_+$ and $\psi_-$ solitons and
was coined  ``twist" (symbolically
$\psi_T$) in \cite{Baer}.
The twist soliton arises both for $h$ greater and smaller than $0.28$.
In the former region the twist obtains
from the $V$-continuation of the $\psi_+$ soliton
while for $h<0.28$, it is  not connected to the $\psi_+$.
(See the solid curve in Fig.\ref{P_of_V}.)
The continuation of the twist in $V$ in the
case $h<0.28$ gives rise to a new branch of
the undamped solutions which has a point of intersection with
the $P=0$ axis, at some $V=V_1$.
A damped traveling wave is
bifurcating off at this value of velocity; see the next subsection.
We are using the same notation $V_1$ in the small-
and large-$h$ case in Fig.\ref{P_of_V} to emphasize the similarity of the
resulting $\gamma(V)$ curves in the two cases  (forthcoming).

Returning to the 
case of large $h$, the entire dashed  curve in Fig.\ref{P_of_V}
corresponds
to symmetric solutions: $\psi(-\xi)={\overline \psi}(\xi)$. 
It turns out that there are also non-symmetric solutions; these were missed in
\cite{Baer}. The real part of a non-symmetric
solution is not even and
imaginary part not odd. In particular, a pair of asymmetric 
solutions arise in
a pitchfork bifurcation of the complex $\psi_{(TT)}$; see the 
dash-dotted offshoot from the dashed curve in Fig.\ref{P_of_V}. 
(The two asymmetric solutions are related by the 
transformation $\psi(\xi) \to {\overline \psi}(-\xi)$; they 
obviously have equal momenta and hence are represented by the 
same  curve.) Continuing the 
asymmetric branch we have the third zero
crossing, at $V=V_3$. When continued to positive $P$,
 the
asymmetric solution acquires the form
of a complex of $\psi_-$ and $\psi_T$ solitons, with the inter-soliton
separation growing as $P$ is increased.
(Note that although the dashed and
dash-dotted curve end at nearby points,  they are {\it not\/}
connected.)
Our numerical analysis shows that branches of damped
solitons do indeed detach at $V_1$, $V_2$ and $V_3$; 
these will be described in
the next two subsections.

\begin{figure}
\includegraphics[ height = 2in, width = 0.5\linewidth]{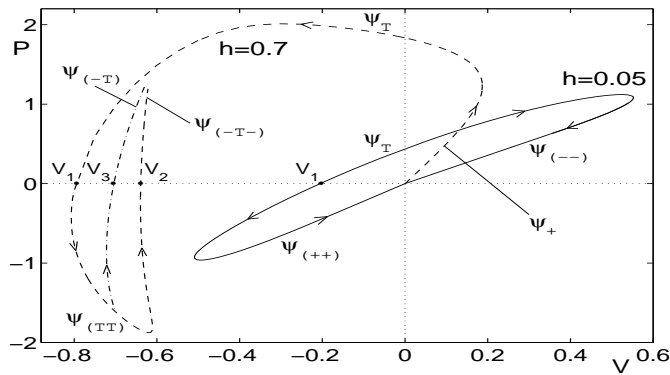}
\caption{\sf The momentum of the undamped traveling wave as
a function of its velocity
(a combined and  advanced version of two diagrams from \cite{Baer}).
 The dashed and dash-dotted curves pertain to
the case of large driving strengths (here exemplified by $h=0.7$).
The starting point $P=V=0$ of the dashed
curve corresponds to the stationary undamped
$\psi_+$ soliton, which then transforms to the
twist, then to a bound state of two twists and then to a complex of a
twist and two $\psi_-$ solitons. (This curve appeared in 
\cite{Baer}.) The dash-dotted offshoot is our new contribution to the 
diagram; it corresponds to an asymmetric solution,
$\psi_{(-T)}$, detaching from the $\psi_{(TT)}$ curve.
The solid curve pertains to
the case of small driving amplitudes
(here $h=0.05$). 
(This curve also appeared in \cite{Baer}.)
The points of its intersection with the $P$-axis
correspond to  stationary twist solitons; continuing each of these
counterclockwise gives rise to a bound state of two $\psi_+$'s, 
while when continued clockwise each twist transforms into
a complex of two $\psi_-$'s. More solution curves can be generated
by the mapping  $V \to -V$, $P \to -P$.
}
\label{P_of_V}
\end{figure}

\subsection{Numerical continuation:
Small driving amplitudes}
\label{Small}

\begin{figure}
\includegraphics[ height = 2in, width = 0.5\linewidth]{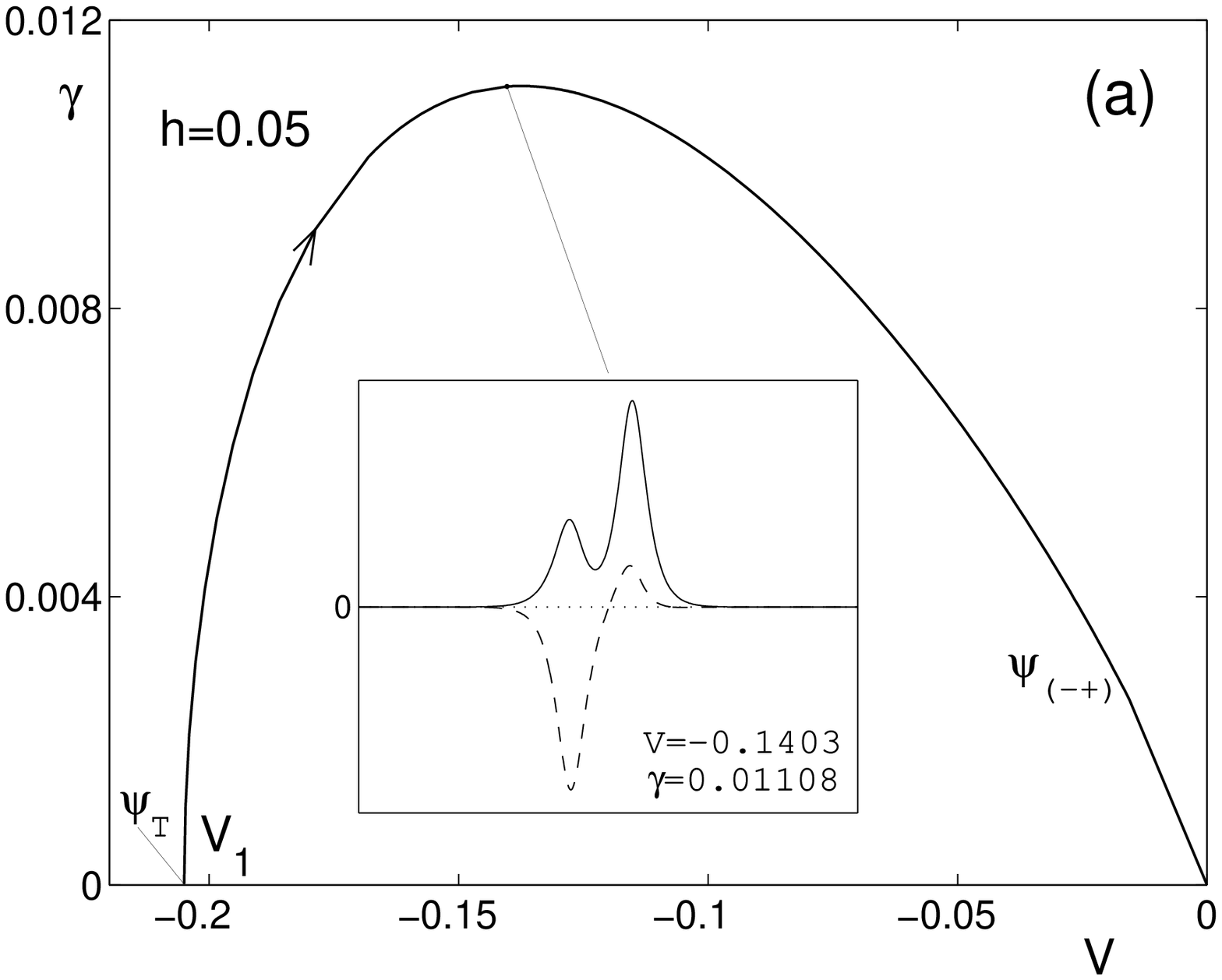}
\includegraphics[ height = 2in, width = 0.5\linewidth]{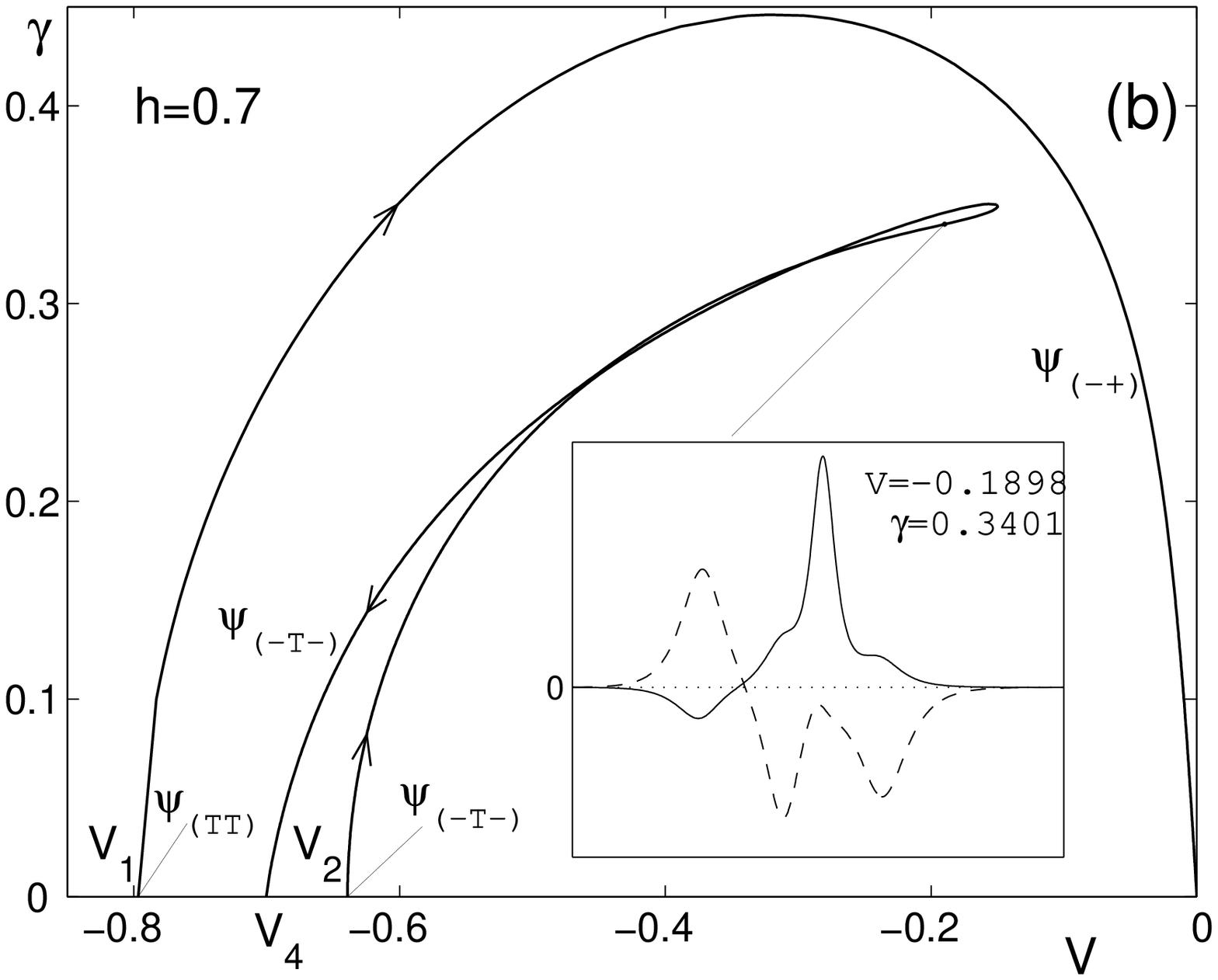}
\includegraphics[ height = 2in, width = 0.5\linewidth]{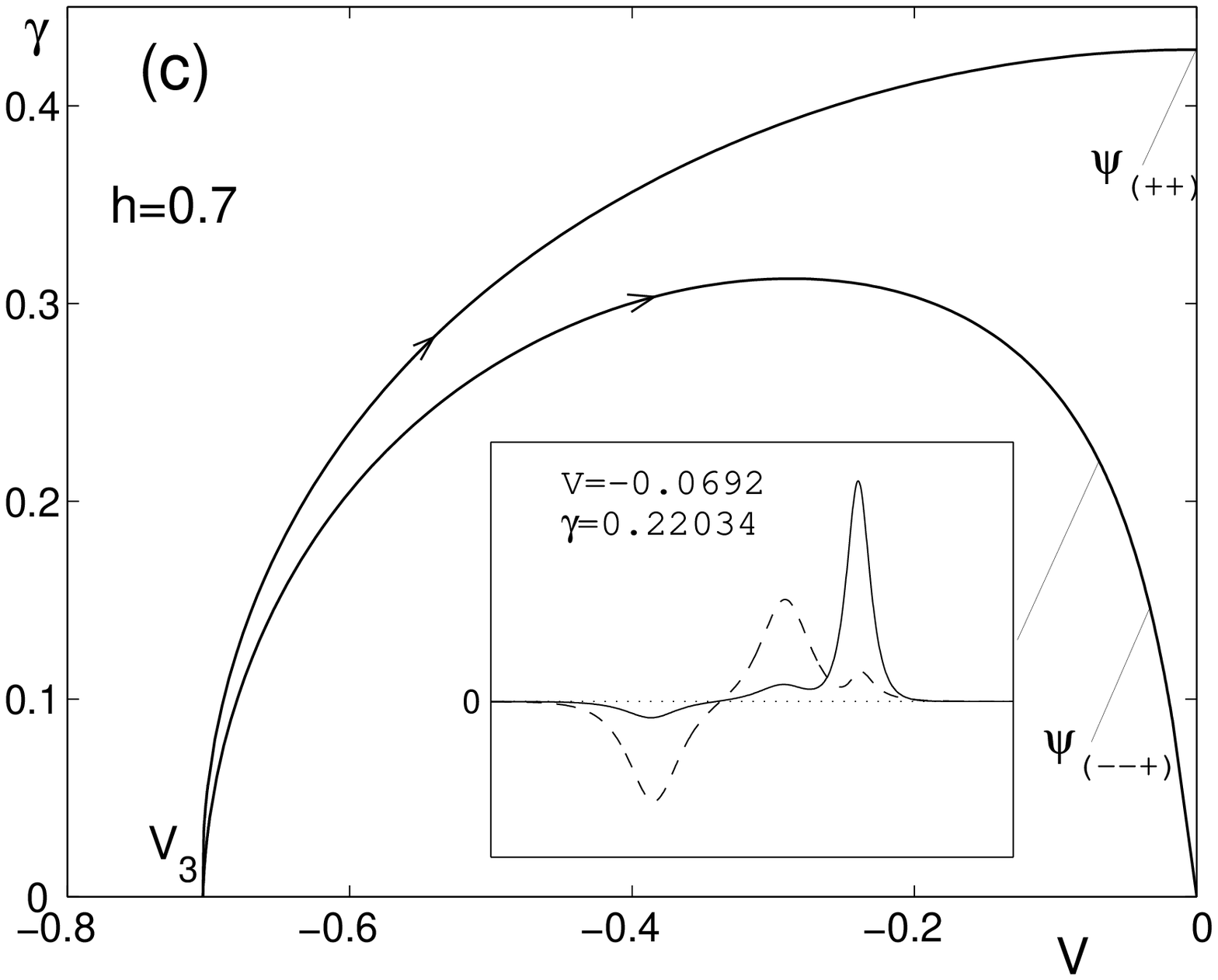}
\caption{\sf Results of the numerical continuation of the undamped
traveling solitons to nonzero $\gamma$.  (a): small $h$; (b),(c): large $h$.
The inset displays a representative
solution at one of the internal
points of the curve. (Solid line: real part; dashed line: imaginary
part.) Each curve shown has a positive-velocity
counterpart which arises by the mirror reflection
$V \to -V$ of the figure.}
\label{gamma_V}
\end{figure}

For small $h$, $h< 0.28$, our continuation departs
from the  twist soliton moving with the velocity
 $V_1$ (the point of intersection of the solid curve
 with the horizontal axis in Fig.\ref{P_of_V}.)
 The real part of this solution is even and imaginary part odd:
$\psi(-x)={\overline \psi}(x)$.
As we continue
to nonzero $\gamma$, this symmetry is lost;
 a typical
profile at the internal points looks like a
non-symmetric complex of the $\psi_-$ and $\psi_+$
and is displayed in the inset to Fig.\ref{gamma_V}(a). The
rest of Fig.\ref{gamma_V}(a)
shows the resulting $\gamma(V)$ dependence.
As $\gamma$  grows, the negative velocity of the traveling
wave decreases in modulus. However the damping cannot
be increased beyond a certain limit value; as we reach it, the
$\gamma(V)$-curve turns down (Fig.\ref{gamma_V}(a)). As $V$
 and $\gamma$ tend to zero, the separation between the
$\psi_-$ and
$\psi_+$ solitons in the complex grows without bounds.

These transformations of the solution
are reflected by the behavior of the
linearized eigenvalues in the eigenvalue problem (\ref{eq25}).
At the point $V=V_1$, $\gamma=0$ of the $\gamma(V)$ curve,
the twist solution has  a quadruplet
of complex eigenvalues $\pm \lambda, \pm {\overline \lambda}$
which dissociates into two pairs
of complex-conjugate eigenvalues $\lambda_1, {\overline \lambda_1}$
and $\lambda_2, {\overline \lambda_2}$ (with ${\rm Re} \, \lambda_1 <0$ and
${\rm Re} \, \lambda_2 >0$) as $\gamma$
deviates from zero. As we move
toward the maximum of the curve, the imaginary parts of $\lambda_1$
and $\lambda_2$ decrease and the four complex eigenvalues move onto
the real axis. At the point of maximum one of the resulting
two positive eigenvalues crosses to the negative real axis, but the other
one persists all the way to $V = -0$
and $\gamma = +0$. Therefore
the
spectrum of eigenvalues on the `downhill' portion of the curve
is a union  of eigenvalues of the $\psi_-$
and $\psi_+$ solitons.
The  conclusion of the eigenvalue
analysis is that the traveling complex whose bifurcation
diagram is exhibited in Fig.\ref{gamma_V}(a), is unstable for all $V$
and $\gamma$.

\subsection{Numerical continuation:
Large driving amplitudes}
\label{Large}

For $h>0.28$  we have three starting points with $P=0$
corresponding to  two intersections of the dashed curve 
and one of the dash-dotted curve
with the horizontal axis in Fig.\ref{P_of_V}.

The $\gamma(V)$ curve emanating out of the point $V_1$
is the top, arc-shaped, curve in Fig.\ref{gamma_V}(b).   For $V=V_1$
 and $\gamma=0$
the solution is  symmetric and its shape
reminds  two strongly
overlapping twists.  The linearized spectrum includes two complex
 quadruplets. As $\gamma$ deviates
from zero,
the symmetry is lost and the solution starts looking 
like an asymmetric complex of two pulses. The two complex
 quadruplets become four complex-conjugate pairs of eigenvalues,
 two with positive and two with negative real parts.
  Two of these pairs (one with $\mbox{Re} \, \lambda>0$
  and one with $\mbox{Re} \, \lambda<0$)
 move on to the real axis.
 After that one positive real eigenvalue crosses to the negative semi-axis,
 while the complex pair with $\mbox{Re} \, \lambda>0$ crosses into
 the $\mbox{Re} \, \lambda<0$ half-plane but then returns to
 $\mbox{Re} \, \lambda>0$.
As $V, \gamma \to 0$,  the separation between
the $\psi_-$ and $\psi_+$ solitons comprising
 this complex increases, and eventually the two
 constituents  diverge to infinities. On the `downhill' portion
 of the curve, the spectrum is a union of the spectra of  the 
 individual $\psi_-$
 and $\psi_+$ solitons; in particular, it includes a positive 
 real eigenvalue and a complex quadruplet. 
 Since there are eigenvalues with ${\rm Re} \, \lambda >0$ for
 all $V$, 
  the entire branch is
 unstable.

The second undamped traveling wave with zero momentum
(point $V_2$ on the bifurcation diagram Fig.\ref{P_of_V})
corresponds to a symmetric
[$\psi(-x)={\overline \psi}(x)$]
complex of two $\psi_-$ and one
twist soliton, symbolically $\psi_{(-T-)}$.
The spectrum includes three complex
quadruplets. As we continue in
$\gamma$ and $V$, the symmetry is lost but the solution still looks
like a complex of three solitons, see the inset to Fig.\ref{gamma_V}(b).
The bottom, spike-shaped, curve in Fig.\ref{gamma_V}(b) depicts
the corresponding $\gamma(V)$ dependence. Unlike
the branch starting at the value $V=V_1$, this solution cannot be continued to
zero velocities. Instead, the $\gamma(V)$ curve turns back
and, as $\gamma$ approaches zero from above, $V$ tends to a negative
value $V_4$, with $|V_4|>|V_2|$. For sufficiently
small $\gamma$ the corresponding solution
consists of  two $\psi_-$ solitons  and
a twist in between, with the inter soliton
separations growing to infinity as
$\gamma \to 0$, $V \to V_4$. The associated eigenvalues perform rather
complicated movements on the complex plane; skipping the details
it suffices to  mention that `unstable' eigenvalues
(real positive or complex with positive real parts) are present for 
all $V$. Hence the entire branch is unstable. 

Finally, the point $V_3$ on the diagram Fig.\ref{P_of_V} represents {\it two\/}
 nonequivalent asymmetric solutions with zero momentum, $\psi_1(\xi)$ and
$\psi_2(\xi)$, with $\psi_2(\xi) \equiv {\overline \psi_1}(-\xi)$. 
Consequently, 
there are {\it two\/} distinct $\gamma(V)$-branches coming out
of this point
(Fig.\ref{gamma_V}(c)). One of these corresponds to 
a complex of two solitons; when continued to $V=0$, it gives rise to
the symmetric complex $\psi_{(++)}$ with nonzero $\gamma$. 
(See the top curve in Fig.\ref{gamma_V}(c)). Continuing the
other asymmetric solution to $V=0$, the corresponding value of $\gamma$
reaches a maximum at $V \sim 0.3$ and then
 tends to zero. (The bottom curve in Fig.\ref{gamma_V}(c)).
For sufficiently small $V$ and $\gamma$ this solution represents
a complex $\psi_{(--+)}$ (shown in the inset to Fig.\ref{gamma_V}(c)).
As $V, \gamma \to 0$, the inter-soliton separation tends to infinity. 
Turning
to the eigenvalues, the start-off solution at the point $V_3$  has two complex
quadruplets and a real positive eigenvalue in its spectrum.
When we continue along the top curve in Fig.\ref{gamma_V}(c),
two complex eigenvalues move on to the positive real axis, so we end up 
with three positive eigenvalues. When we continue 
along the bottom curve, 
the movements of the eigenvalues are more involved
but some of them always remain in the `unstable' half-plane,
${\rm Re} \, \lambda >0$. The upshot
of the eigenvalue analysis is that both curves represent only unstable
solutions.

\section{Consistency of the two approaches}
\label{consistency}

\begin{figure}
\includegraphics[ height = 2in, width = 0.5\linewidth]{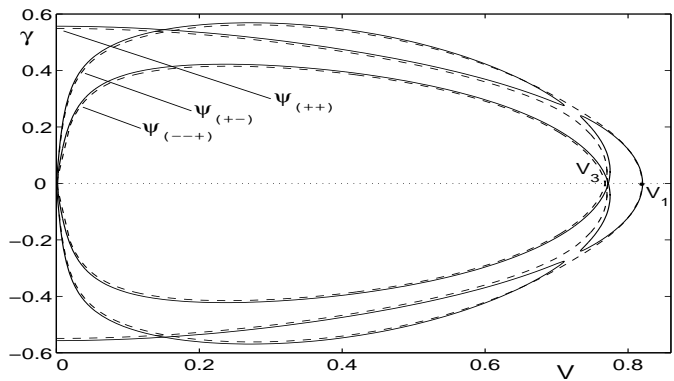}
\caption{\sf The comparison of the bifurcation diagrams
for $h=0.82$ and $h=0.8275$. Solid curve: $h=0.8275$;
dashed curve: $h=0.82$. Although the  
points of intersection of the dashed and solid curves 
with the $V$-axis do not coincide, i.e.
$V_3(0.8275) \neq V_3(0.82)$
and  $V_1(0.8275) \neq V_1(0.82)$, they are quite close
to each other
and hence we mark them as a single point ($V_3$ and $V_1$,
respectively.) 
The branches coming out of the point $V_2$ on the $V$-axis
are omitted for visual clarity.
Note the shape of the dashed
and solid curves near $V \sim 0.7$, $\gamma \sim 0.3$,
characteristic of phase portraits of 2D dynamical systems
 in the neighborhood of
saddle points.
 }
\label{cusps}
\end{figure}

To complete our classification
of damped traveling solitons, we need to comment on what may seem to be
an inconsistency between results
obtained within the above two complementary approaches.
The solution representing  the well-separated $\psi_+$ and $\psi_-$
solitons
 reported in sections \ref{nonzero_gamma}
 and  \ref{nonzero_V}, can be reached
 by continuing both off the $(\gamma=0)$- and  $(V=0)$-axes.
(This branch  connecting to
the origin on the $(V, \gamma)$-plane appears
both in Figs. \ref{conti}(a)
and  \ref{gamma_V}(b).) 
Although such a curve should
obviously not depend
on the starting point of the continuation, one notices that the
$\psi_{(+-)}$ branches `flowing into the origin'  in Figs. \ref{conti}(a)
and  \ref{gamma_V}(b) behave differently when traced backward
(i.e. away from $V=\gamma=0$). While the curve in Fig.\ref{conti}(a)
intersects the $\gamma$-axis, its counterpart in Fig.\ref{gamma_V}(b)
crosses the other, $V$-, axis.
(Here the reader should not be confused 
by the fact that the 
$\psi_{(+-)}$ branch
in Fig.\ref{conti}(a) is shown for positive and its counterpart in 
Fig.\ref{gamma_V}(b) for negative values of $V$. In  view of the  
$\xi \to -\xi$, $V \to -V$ invariance of equation (\ref{eq8}),
to each $\gamma$ there correspond {\it two\/} traveling waves, one
with  positive and the other one with negative
value of $V$. Therefore, one should
 mirror-reflect Fig.\ref{gamma_V}(b)
prior to comparing it to Fig.\ref{conti}(a). This reflection 
maps the solution $\psi_{(-+)}$ of Fig.\ref{gamma_V}(b) to
the $\psi_{(+-)}$ of Fig.\ref{conti}(a).) 

To resolve the  paradox, one needs to note that the two figures
correspond to different values of $h$,
Fig.\ref{conti}(a) to $h=0.8353$ and Fig.\ref{gamma_V}(b) to $h=0.7$. 
It turns out that a qualitative change of
behavior 
 occurs for $h$ somewhere between these
values, more precisely  between $0.82$ and $0.8275$:
For $h=0.82$ and smaller
(in particular, for $h=0.7$) the $\gamma(V)$ curve has the form of
an arc shown in 
 Fig.\ref{gamma_V}(b)
(i.e. it crosses the $V$-axis as $V$ is increased) while   
 for $h=0.8275$ and greater, 
the curve is already loop-shaped and does not reach to $\gamma=0$.
This change of behavior, 
accounting for the above 
`inconsistency', is illustrated by Fig.\ref{cusps} which
compares the $\gamma(V)$ dependences for $h=0.82$ and $h=0.8275$.
Fig.\ref{cusps} also serves to illustrate the different outcomes of
 the continuation of the complex $\psi_{(++)}$ 
for  $h=0.8353$ and 
smaller $h$. (We note that for $h=0.8353$
the continuation of the motionless $\psi_{(++)}$ produces 
a pair of
infinitely separated solitons $\psi_+$ and $\psi_-$
 [Fig.\ref{conti}(a)] while for $h=0.7$, 
the curve departing from the same type of starting point
 [i.e. from $\psi_{(++)}$] ends up at the undamped
asymmetric solution traveling with nonzero velocity $V_3$
[Fig.\ref{gamma_V}(c)].)

The above differences in behavior result from the presence
of a saddle point on the $(\gamma,V)$-plane,
in the gap between the two lobes of the solid curve in Fig.\ref{cusps}.
Indeed, the dashed and solid curves can be seen as sections
   of the surface $h=h(\gamma,V)$
 by the horizontal planes $h=0.82$ and $h=0.8275$, respectively.
 The gap in the upper solid curve is then accounted for by letting
$h= h_0+ x^2-y^2$ in the vicinity of the gap.
Here the constant $h_0$ lies somewhere between $0.82$
and  $0.8275$, and  $(x, y)$ is a pair of 
suitably chosen coordinates on the $(\gamma,V)$-plane. 

\section{Conclusions}
\label{conclusions}

One of the conclusions of this work is that by
grouping into complexes, solitons (or, equivalently, solitary pulses)  can 
adjust their total momentum to zero. By doing so they can
travel with nonzero speed in the presence of damping ---
without violating the momentum decay law, ${\dot P}=- \gamma P$. 
Two identical solitons traveling at the same speed in the same
direction have equal momenta; therefore,
in order to arrange for $P=0$ the traveling complex inevitably has
to include solitons of different varieties (i.e. both $\psi_+$'s
and $\psi_-$'s.) Consequently, the real and imaginary parts
of the traveling complex will always be represented by {\it
asymmetric\/} functions of $\xi=x-Vt$. 

Although the possibility of non-decelerated motion
may be out of line with the common
 perception of the soliton dynamics in 
 weakly damped Hamiltonian equations, moving pulses are not unknown 
in {\it strongly\/} dissipative systems. A  suitable example
is given by the complex Ginzburg-Landau equation. Asymmetric 
Ginzburg-Landau pulses, uniformly traveling with nonzero 
velocities, were reported in \cite{Afanasjev}.

All moving solutions that we have found in this paper,
 turned out to be unstable. 
 This instability admits a simple qualitative explanation --- at least,
  for small dampings.
 In the undamped situation, the $\psi_-$ solitons 
 are unstable   when traveling with small velocities while the
 $\psi_+$'s become unstable when moving sufficiently
 fast \cite{Baer}. In the presence of dissipation
  the traveling 
 wave has to include solitons of  both varieties; 
 on the other hand, the eigenvalues corresponding to small
 nonzero $\gamma$ should remain close to their 
 ($\gamma=0$)-counterparts. Therefore
  the  spectrum of the traveling complex
 will `inherit' unstable eigenvalues of either $\psi_-$ (for small
 velocities) or of the $\psi_+$ (for large velocities).

 Thus,
despite the fact that the parametric driver can sustain
the uniform motion of a damped soliton,
an additional  agent (such as, possibly,
the diffusion and/or a nonlinear damping term)
is required to make this motion stable.
Here it is appropriate to refer, again, to the complex 
Ginzburg-Landau equation. Stable Ginzburg-Landau pulses arise as a
result
of a delicate balance of the whole series of terms, including
dispersion, cubic and quintic nonlinearity, diffusion, cubic gain and
linear and quintic nonlinear damping
\cite{stableGL,Afanasjev,Akhmediev}. 
In a similar way, the gain/loss and 
spreading/steepening balances of the damped-driven traveling 
solitons could be restored by adding one or several missing
agents.

\acknowledgments

We thank Nora Alexeeva for her advice on numerics.
The first author's (I.B.'s) work was supported
 by the NRF
 of South Africa under grant No.2053723;
 by the Johnson Bequest Fund and the URC of the University of Cape Town.
The second author (E.Z.) was  supported by the Russian
Foundation for Fundamental Research under grant No.0301-00657;
by the Visiting Lecturer's Fund of UCT, 
and by an NRF travel grant No.2060193.

\end{document}